\documentclass[aps,pre,preprint,groupedaddress,nofootinbib,showpacs,tightenlines]{revtex4}

\usepackage{amsfonts,amsmath,amssymb,bm}
\usepackage{graphicx}
\usepackage{dcolumn}

\begin{document}

\title{On Application of Fractal Analysis to Cranial Sutures}

\author{Andrzej Z. G\'orski}%
\email{Andrzej.Gorski@ifj.edu.pl}
\affiliation{Institute of Nuclear Physics, %
Polish Academy of Sciences,\\
Cracow, Poland}
\author{Janusz Skrzat}
\email{jskrzat@poczta.onet.pl}
\affiliation{%
Dept.~of Anatomy, Collegium Medicum, Jagiellonian University,\\
Cracow, Poland}

\date{\today}

\begin{abstract}
 Fractal exponents ($d$) for human cranial sutures are calculated 
using the box  counting method. The results were found around $d = 1.5$
(within the range $1.3\div 1.7$), supporting 
the random walk model for the suture formation process. 
However, the calculated dispersion above the estimated accuracy suggests
that other mechanisms are also present. 
Similar numbers were obtained for both the sagittal and coronal sutures, 
with the coronal sutures displaying a better scaling. 
 Our results are compared with estimations published by other 
authors. 
\end{abstract}


\maketitle

\section{Introduction}

 Geometry of a physical object is often characterized by its fractal
dimension\cite{Renyi56,Renyi70,MANDELBROTa}.
 This is an adequate description of self-similar sets widely
used in various applications, ranging from high energy 
\cite{FRACinEP} and condensed matter physics \cite{FRACinCM} to
astrophysics \cite{FRACinAP} and econophysics \cite{azgDAXb}. 
 Fractal properties of an object impose strong conditions upon its 
structure and, in particular, 
upon a scaling symmetry that is difficult to overestimate 
in physical applications. It is also a method for quantifying and 
comparing the spatial complexity of real objects, characteristic
of the image as a function of scale. This seems to be 
especially well suited for morphological analysis. 
 These properties can be used for classification, as well as 
they should be taken into account in models of morphogenesis. 

 In this paper, we estimate the fractal dimension of the human cranial
sutures. 
In recent years there were several attempts in this direction
\cite{Hartwig91,Skrzat00,Skrzat03a,Jack03,Lynnerup03}. 
However, to determine the fractal dimension the authors have used 
an automated commercial software and the obtained values were 
highly dispersed, the results ranging from $1.0$ up to over $1.7$. 
In some cases, rather unphysical results in the range 
$0.0 \div 1.0$ were also reported \cite{Lynnerup03} --- 
the cranial suture is a curve and its fractal dimension must 
be within the range $1.0 \div 2.0$. 
Similar confusions occur also in other areas where a fractal 
dimension was estimated \cite{blad}.

 In our analysis, we point out the subtleties 
in determining the fractal dimension that are immanent for such 
numerical estimates, especially for a relatively limited number 
of data points, as is in our case. 
 In particular, the crucial point in extracting the fractal dimension 
from a typical log--log plot is the proper choice of points 
that are used to get the linear fit. 
These points should be taken from the scaling region
in order to avoid boundary effects
that are usually stronger for not very large data sets. 
Hence, this task should not be left to the automated procedures. 
Instead, one should perform a careful analysis which 
points are to be choosen. 
It will be shown that the standard deviations 
in the least square fits for two different choices of fitted 
points can be very small but, in spite of that, 
both fits lead to considerably different results. 
Finally, we try to estimate the accuracy of our calculations.

\section{Collecting the data}

The skulls are stored at the Department of Anatomy of Collegium 
Medicum of Jagiellonian University. 
They belong to adult individuals of different sex and derived from 
different populations. 
The investigated cranial set presents considerable morphological 
variation caused by diversified ethnicity and historical dating. 
Hence, we can expect relatively large standard deviation of 
calculated fractal exponents for different skulls.

\begin{figure}
  \includegraphics[width=8.5cm,angle=0]{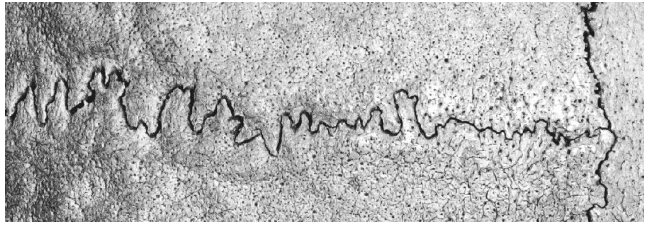} 
  \caption{Sagittal suture 03s.}
\label{fig:szew}
\end{figure}

The numerical data have been collected from forty cranial sutures
of two types: twenty sagittal and twenty coronal ones. 
Seventeen of them are from the same skulls and in the last three 
cases the two types of sutures come from different skulls 
(to have sutures of better quality). 
All examined sutures were non-obliterated and their contours were well 
visible.
The segments of the coronal and sagittal sutures, which were visible 
in superior aspect of the cranial vault, were photographed using 
the digital camera.  
The images have sufficient resolution, above $0.1$~mm; 
a sample image for the sagittal suture $03s$ is displayed 
in Fig.~\ref{fig:szew}. 
From these images, using Micrografx Picture Publisher software 
we extracted one-pixel line being the sutural contour. 
The pixels were converted into the numbers (2-dimensional coordinates)
that were directly used to calculate the fractal dimensions. 

We estimate the final accuracy (taking into account details of the suture
border) within the range between $0.1$ and $0.5$~mm,
similarly as the other authors \cite{Hartwig91,Jack03}, while
the size of the whole sample is of order $100$~mm. 
\textit{i.e.}, up to about three orders of magnitude.
The resulting maximum number of fitted points in the log--log plot
should be less than eight. 
This implies that the expected scaling cannot extend above 
the range of two orders of magnitude. 

We have assumed that the sutures are basically 2-dimensional objects, 
as the skull's surface is relatively smooth and the fractal dimensions 
of sets should be invariant with respect to their smooth 
transformations. 
Hence, the numerical data can be viewed as a local projection
of sutures to the $xy$-plain. 
As a result, the initial numerical data are sets of the
coordinates ($x,y$). 
The number of data points ($n_{tot}$) in each set varies within the 
interval $9~000 \div 29~000$, with the average of above $16~000$. 
For 2-dimensional sets, this is not a very large number and 
determination of the best fit in the log--log plot should be 
performed very carefully.

\section{The Box Counting Algorithm}

 To determine the fractal properties of our data we use the standard 
box counting algorithm for sets embedded in 2-dimensional plain. 
The generalized Renyi exponents, $d_q$, are defined by \cite{Renyi70}
\begin{equation}
\label{dqDEF}
d_q = \frac{1}{1-q} 
\lim_{N\to\infty}
\frac{\ln \sum_i p_i^q(N)}{\ln N} = 
\lim_{N\to\infty} \frac{\ln Y_q(N)}{\ln N}
 \ ,
\end{equation}
where $N$ is the number of successive division of a plain into squares
(``boxes''), $p_i(N)$ denote the fractions of the data contained in the 
$i$th box of a given division ($N$), and $Y_q(N)$ is defined by the 
second equality. 
The box size, $\epsilon \sim 1/N$. 
The value of $d_q$ can be extracted from the log--log plot of the value
$Y_q(N)$ \textit{vs} $N$. 

Clearly, for any data set, whether fractal or not, we get the first
point ($N=1$) in the log--log plot with coordinates 
$(0,0)$ ($p_1(1)=1$). 
In addition, for any ``practical'' (finite) set with large enough $N$,
the value of $Y_q(N) \to n_{tot} ~~(N\to\infty)$, 
where $n_{tot}$ is the total number of data points 
(the division becomes so fine that each data point occupies
a separate box). 
In both these regions (too small and too large $N$) 
the boundary effects become dominant and they
are not credible to get an estimate for the fractal dimensions. 
Hence, one should have a large enough
data set to obtain reliable results (sufficient number of fitted points), 
though the standard fit measures like
$\chi^2$ or Pearson's $r$--coefficient, may look satisfying.
This is the main reason why 
the whole procedure is rather subtle and can give misleading
results. Hence, we have used our dedicated box counting code instead of 
a commercial software to be able to control the intermediate
steps. 

\begin{figure}
  \includegraphics[width=8.5cm,angle=0]{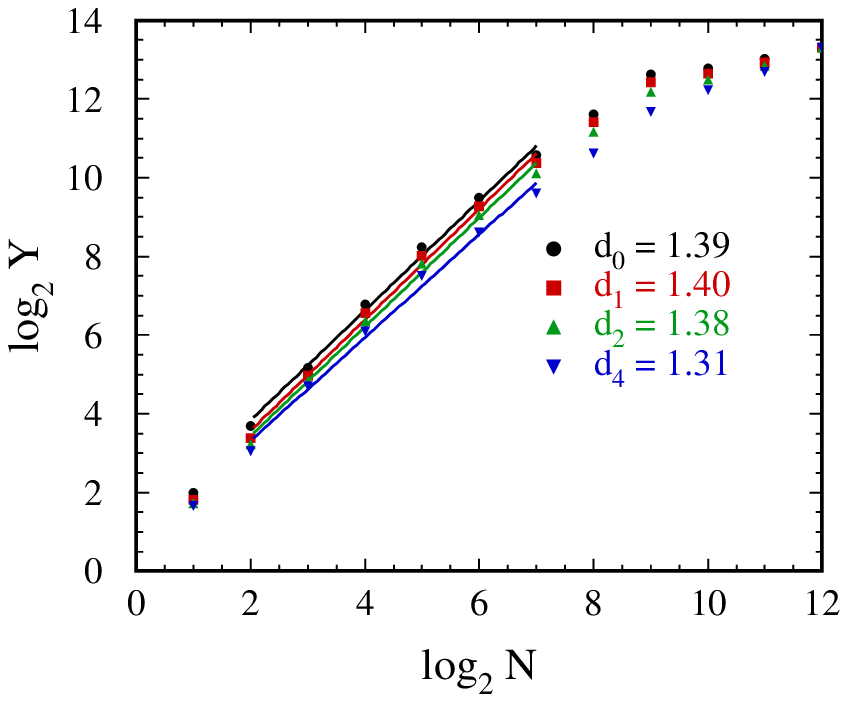} 
  \caption{Example of the log-log plot for the sagittal suture 03s used to extract Renyi exponents in case of 6-point fits. Fits for $d_q$ with q = 0, 1, 2, and 4 are displayed.}
\label{fig:fig1}
\end{figure}

In our case, the number of points in the log--log plot that 
is within the reasonable scaling range (linear scaling)
equals $5\div 7$ ---
for all plots we repeat calculations for all these three cases
(5, 6 and 7 points) to see differences. 
The first point we take into account is $N=3$ (the box size is 
$\frac{1}{8}$ \textit{i.e.}, of order of $1$~cm). 
The upper limit is restricted by the sample size. 
The corresponding log--log plots are of definitely better quality 
for the coronal sutures. 
In Fig.~\ref{fig:fig1} we show, as an example, the
plot for the sagittal suture $03s$, corresponding to the first
raw in Table \ref{tab:table1} (for the coronal sutures 
the fits are better). 
As can be seen from the plot ,
the saturation of scaling is clearly visible for $N=10$. 
Taking into account the points from $N=3$ up to $N=9$, we have found that 
the results considerably deviate from the case when 
the scaling exponents are calculated up to the $N=8$ and $N=7$ 
subdivisions. This is consistent with the fact that, in the subdivision 
with the factor $2^9=512$, the details well below $0.5$~mm became 
important. 
The results for the latter two cases are given in 
Tables~\ref{tab:table1}~and~\ref{tab:table2}.

The best linear fit is reached in the case of 6 points.
The $\chi^2$ parameter (a sum of the squared deviations from the fit) 
for $d_0$ was below $0.05$ in most cases. Typically, it was much better 
for the coronal sutures (on average $0.004$ and $0.005$ for 6-point 
and 5-point fits, respectively, leading to the expected ``theoretical''
accuracy of $d_q$ below $0.01$). 
For the sagittal sutures, the corresponding values are $0.05$ and $0.02$,
which gives uncertainty for $d_0$ below $0.02$. 
However, as was stated earlier, one should be aware that a high quality 
of the linear fit does not guarantee a high accuracy for the resulting 
real fractal exponent:

\textit{(i)} a proper choice of the points in the
log--log plot is very important and is rather difficult to implement 
for automated algorithms;

\textit{(ii)} for some pathological data sets one can obtain misleading 
results due to their pseudofractal nature \cite{azgPSEUDOFRAC}, in spite
of excellent numerical scaling;

\textit{(iii)} the noise level in the data is unknown and we cannot 
estimate the real error bars for the point in the log--log plot;

To get reliable results we compare the output of our code
for the $d_q$--exponents
for three adjacent choices of linear fits and for four values of the
parameter $q = 0, 1, 2$, and $4$. 
This has been done to see the resulting differences in the values
of the fractal exponents and to check for possible 
multifractal and pseudofractal effects. 
Usually this method provides the fractal exponent with the accuracy 
of about a few percent.

\begingroup
\squeezetable 
\begin{table}
  \caption{\label{tab:table1} The Renyi exponents for the sagittal sutures.}
  \begin{ruledtabular}
    \begin{tabular}{cdddd}
      suture & d_0 & d_1 & d_2 & \chi^2 \\
      \hline
      03s & 1.39 & 1.40 & 1.38 & 0.04 \\
          & 1.46 & 1.48 & 1.46 & 0.01 \\
      04s & 1.53 & 1.53 & 1.51 & 0.05 \\
          & 1.62 & 1.61 & 1.58 & 0.03 \\
      06s & 1.54 & 1.55 & 1.55 & 0.05 \\
          & 1.62 & 1.64 & 1.63 & 0.01 \\
      10s & 1.47 & 1.48 & 1.46 & 0.07 \\
          & 1.56 & 1.57 & 1.56 & 0.03 \\
      14s & 1.56 & 1.56 & 1.56 & 0.07 \\
          & 1.66 & 1.66 & 1.63 & 0.02 \\
      16s & 1.57 & 1.57 & 1.55 & 0.04 \\
          & 1.63 & 1.64 & 1.63 & 0.01 \\
      17s & 1.51 & 1.50 & 1.46 & 0.08 \\
          & 1.61 & 1.59 & 1.55 & 0.04 \\
      18s & 1.57 & 1.59 & 1.57 & 0.04 \\
          & 1.65 & 1.67 & 1.66 & 0.01 \\
      19s & 1.44 & 1.47 & 1.47 & 0.03 \\
          & 1.50 & 1.55 & 1.54 & 0.01 \\
      20s & 1.58 & 1.61 & 1.59 & 0.06 \\
          & 1.67 & 1.70 & 1.69 & 0.02 \\
      22s & 1.59 & 1.61 & 1.59 & 0.03 \\
          & 1.65 & 1.68 & 1.67 & 0.01 \\
      23s & 1.43 & 1.46 & 1.46 & 0.04 \\
          & 1.51 & 1.55 & 1.54 & 0.01 \\
      24s & 1.56 & 1.55 & 1.53 & 0.07 \\
          & 1.66 & 1.65 & 1.62 & 0.02 \\
      25s & 1.55 & 1.56 & 1.54 & 0.08 \\
          & 1.65 & 1.65 & 1.63 & 0.03 \\
      27s & 1.50 & 1.52 & 1.50 & 0.04 \\
          & 1.57 & 1.59 & 1.58 & 0.01 \\
      28s & 1.45 & 1.47 & 1.44 & 0.04 \\
          & 1.53 & 1.55 & 1.53 & 0.01 \\
      30s & 1.50 & 1.52 & 1.50 & 0.05 \\
          & 1.58 & 1.61 & 1.59 & 0.01 \\
      \hline
      01s & 1.40 & 1.43 & 1.40 & 0.04 \\
          & 1.48 & 1.51 & 1.49 & 0.01 \\
      21s & 1.48 & 1.51 & 1.49 & 0.05 \\
          & 1.57 & 1.60 & 1.59 & 0.02 \\
      26s & 1.48 & 1.50 & 1.49 & 0.04 \\
          & 1.55 & 1.57 & 1.56 & 0.02 \\
    \end{tabular}
  \end{ruledtabular}
\end{table}

\begin{table}
  \caption{\label{tab:table2} The Renyi exponents for the coronal sutures.}
  \begin{ruledtabular}
    \begin{tabular}{cdddd}
      suture & d_0 & d_1 & d_2 & \chi^2 \\
      \hline
      03c & 1.44 & 1.45 & 1.44 & 0.006 \\
          & 1.47 & 1.48 & 1.47 & 0.002 \\
      04c & 1.36 & 1.37 & 1.36 & 0.001 \\
          & 1.38 & 1.38 & 1.38 & <0.001 \\
      06c & 1.48 & 1.53 & 1.54 & 0.005 \\
          & 1.49 & 1.55 & 1.56 & 0.001 \\
      10c & 1.35 & 1.35 & 1.35 & 0.004 \\
          & 1.37 & 1.37 & 1.36 & 0.003 \\
      14c & 1.48 & 1.51 & 1.53 & 0.006 \\
          & 1.51 & 1.55 & 1.58 & <0.001 \\
      16c & 1.62 & 1.63 & 1.61 & 0.008 \\
          & 1.65 & 1.68 & 1.66 & <0.001 \\
      17c & 1.51 & 1.53 & 1.52 & 0.003 \\
          & 1.52 & 1.55 & 1.55 & 0.003 \\
      18c & 1.30 & 1.30 & 1.30 & 0.003 \\
          & 1.30 & 1.29 & 1.29 & 0.004 \\
      19c & 1.28 & 1.29 & 1.28 & 0.006 \\
          & 1.35 & 1.35 & 1.34 & 0.009 \\
      20c & 1.49 & 1.51 & 1.51 & 0.003 \\
          & 1.49 & 1.51 & 1.52 & 0.004 \\
      22c & 1.58 & 1.61 & 1.61 & 0.012 \\
          & 1.62 & 1.66 & 1.67 & 0.003 \\
      23c & 1.45 & 1.47 & 1.47 & 0.002 \\
          & 1.46 & 1.48 & 1.48 & 0.001 \\
      24c & 1.57 & 1.62 & 1.60 & 0.010 \\
          & 1.61 & 1.67 & 1.66 & <0.001 \\
      25c & 1.60 & 1.65 & 1.67 & 0.005 \\
          & 1.63 & 1.69 & 1.71 & 0.001 \\
      27c & 1.34 & 1.37 & 1.37 & 0.013 \\
          & 1.33 & 1.38 & 1.37 & 0.017 \\
      28c & 1.44 & 1.45 & 1.43 & 0.004 \\
          & 1.46 & 1.46 & 1.43 & 0.002 \\
      30c & 1.44 & 1.51 & 1.52 & 0.002 \\
          & 1.45 & 1.53 & 1.55 & 0.002 \\
      \hline
      07c & 1.48 & 1.51 & 1.52 & 0.010 \\
          & 1.51 & 1.56 & 1.57 & 0.030 \\
      11c & 1.34 & 1.39 & 1.39 & <0.001 \\
          & 1.35 & 1.39 & 1.39 & <0.001 \\
      13c & 1.44 & 1.52 & 1.53 & 0.006 \\
          & 1.43 & 1.52 & 1.54 & 0.006 \\
    \end{tabular}
  \end{ruledtabular}
\end{table}
\endgroup

\section{Numerical Results}

 In our case, the data series are of the average length ($n_{tot}$)
of $16~400$. For such samples, 
the crude estimate of the standard Gaussian error 
is of the order of $1/\sqrt{n_{tot}} \sim 1\%$ 
\textit{i.e.} about $\pm 0.01$ for the estimated value of $d_q$. 
This is quite close to the standard deviations for the linear fits. 
However, taking into account the problem with a choice of the points 
to be fitted in the log--log plot, the accuracy in this
type of calculations can be roughly \textit{estimated} to be
about $\pm 0.05$, \textit{i.e.}, about a few percent. 
The accuracy of those calculations is being depreciated with growing 
value of the parameter $q$. While our results 
for $q = 0, 1, 2$ (the capacity, information, and correlation dimension)
are the same within $\pm 0.02$, the calculated values of $d_q$, 
for $q=4$, differ from the lower $q$'s 
(the difference equals about $0.1$ and less). 
Hence, only the former values of the least square fit are presented
in Table~\ref{tab:table1}~and~\ref{tab:table2}.

In the first column, the suture code is given. The number denotes the 
skull and the labels ($s,c$) denote the suture type (sagittal,
coronal). 
In all the cases the points fitted in the log--log plot were 
located in the interval $N=3\div 8$ (eq. (\ref{dqDEF})),
where the point $(0,0)$ is assumed to be the first point 
and the successive subdivisions differ, as usually, by factor two. 
The corresponding five-point fit is also given in all cases, below 
the the 6-point fit. This is to show the differences for the two 
neighboring fits. The 7-point fit was also done, however 
the results are not given in the Tables. 
In the last column, the estimate for the $\chi^2$ value is given. 
These calculations clearly shows that:

\textit{(A)} For the both, coronal and sagittal sutures, the average
differences $\vert d_0-d_1\vert$ and 
$\vert d_1-d_2\vert$ are equal to about $0.01$, which is
negligible in comparison to the estimated accuracy; the same result can 
be found for 5, 6 and 7-point fits; this suggests a good 
\textit{monofractal scaling} within the investigated range 
(two orders of magnitude);

\textit{(B)} For the difference $\vert d_2-d_4\vert$, taking into 
account all the calculated fits, we get the difference of $0.02$ 
for coronal and of $0.07$ for sagittal sutures; 
\textit{i.e.}, the coronal sutures are 
less ``noisy''; this is partly due to the data series that are by
about $15$\% longer in this case; 
however, it is possible that this effect is also connected with 
the better scaling of the coronal data series;

\textit{(C)} The 5-point fit gives a systematic difference in comparison
with the 6-point fit, on average about $+0.02$ for the coronal and 
$+0.08$ for the sagittal sutures; this suggests that we are already 
approaching the boundary of scaling and this is another indication 
of the superior scaling for the coronal sutures; 

\textit{(D)} The average values for 5-point and 6-point fits to 
$d_0$, $d_1$, and $d_2$ and the corresponding standard 
deviations are given in Table \ref{tab:table3}; 
the values vary in the range $1.38 \div 1.70$
(sagittal sutures, Table~\ref{tab:table1}) and 
$1.29 \div 1.71$ (coronal sutures, Table~\ref{tab:table2}), 
the average values are very close to $1.5$ with 
the standard deviation of about $0.1$; 

\textit{(E)} The diffrences between different sutures are larger than 
the expected accuracy;

\textit{(F)} the results for the coronal sutures are a few percent lower
than these for the sagittal ones; as the difference is comparable with
the estimated error, it can be caused by the bigger noise component;

\begin{table}
  \caption{\label{tab:table3} Average values of Renyi exponents.}
  \begin{ruledtabular}
    \begin{tabular}{lcddd}
      \ suture type & fit & d_0 & d_1 & d_2 \\
      \hline
      sagittal & 6-point & 1.51 & 1.52 & 1.50 \\
      sagittal & 5-point & 1.59 & 1.60 & 1.59 \\
      total average  &  & 1.55 & 1.56 & 1.55 \\
      \hline
      coronal & 6-point & 1.45 & 1.48 & 1.48 \\
      coronal & 5-point & 1.47 & 1.50 & 1.51 \\
      total average  &  & 1.46 & 1.49 & 1.50 \\
    \end{tabular}
  \end{ruledtabular}
\end{table}

\section{Summary and Conclusions}

 A detailed analysis of fractal properties of the coronal
and sagittal sutures has been performed by applying the box counting 
algorithm for the data series consisting of about $17~000$ points. 
The final values of the Renyi exponents are equal to 
about $1.55$ for sagittal and $1.47$ for coronal sutures. 
The least square fit ($\chi^2$) for all fits is rather good 
implying the error below $\pm 0.05$ for the $d_q$ values,
and even lower for the coronal sutures. 

We point out that the differences caused by the choice of points to be
fitted in the log--log plot lead to higher differences, about $0.10$ 
for the optimal choices: $N=3\div 8$ and $N=3\div 7$. 
In our opinion, this is the main source of considerable differences in 
the results published by other authors (see \textit{e.g}, Refs.
\cite{Hartwig91,Skrzat00,Skrzat03a,Jack03,Lynnerup03}).
This applies to other estimates of fractal properties, especially
in those cases where the number of data points is relatively modest. 

The average value of the fractal exponent for all the sutures taken into
account is, within the estimated accuracy, close to $1.5$, 
the exact value for the \textit{Brownian random walk}. 
This suggests that future models of the suture formation should 
have the random-walk model as the basic ingredient. 
For various skulls under investigation, the results differ 
by up to $\pm 0.2$. This indicates that the random walk 
mechanism, though essential, is not sufficient to explain
the suture formation and further investigation is necessary.

\bibliography{gorski}

\end{document}